\documentclass[preprint]{emulateapj}
\usepackage{graphicx}
\usepackage{ulem}
\usepackage{natbib}
\begin{document}

%\title{Predicted 
\title{Spectral Energy Distributions of Accreting Protoplanets}

\author{J.A. Eisner\altaffilmark{1}}
\affil{Steward Observatory, The University of Arizona, 933 N. Cherry
  Ave, Tucson, AZ 85721}

\email{jeisner@email.arizona.edu}

%\and
%\author{P.J. Armitage}
%\affil{JILA, University of Colorado and NIST, and
%Department of Astrophysical and Planetary Sciences, University of
%Colorado, Boulder, CO 80309}

\altaffiltext{1}{Visiting Fellow, JILA, University of Colorado and
  NIST, Boulder, CO 80309}

%\keywords{stars:pre-main sequence---stars:circumstellar 
%matter}

%\slugcomment{Draft of {\bf \today}}

\begin{abstract}
Planets are often invoked as the cause of inferred gaps or inner clearings in
transition disks.  These putative planets would interact with the remnant
circumstellar disk, accreting gas and generating substantial luminosity.
Here I explore the expected appearance of accreting protoplanets at a
range of evolutionary states.  I compare synthetic spectral energy
distributions with the handful of claimed detections of
substellar-mass companions in transition disks.  While observed
fluxes of candidate companions are generally compatible with accreting
protoplanets, challenges remain in reconciling the extended structure
inferred in observed objects with the compact emission expected from
protoplanets or circumplanetary disks.
I argue that a large fraction of transition disks should
harbor bright protoplanets, and that more may be detected as larger
telescopes open up additional parameter space.
\end{abstract}


\section{INTRODUCTION \label{sec:intro}}

Transition disks exhibit spectral energy distributions (SEDs) similar
to classical T Tauri or Herbig Ae/Be stars, but with a deficit of
emission at near-IR wavelengths \citep[e.g.,][]{STROM+89,CALVET+02}.  
This lack of near-IR excess emission suggests inner clearings. 
Direct imaging has now confirmed the existence of
such inner holes, and indicated sharp edges between the optically
thick outer disks and the cleared inner regions
\citep[e.g.,][]{HUGHES+07,ANDREWS+11}.

Dynamical interactions of a massive planet and disk can cause sharp
edges like those seen in transition disks \citep[e.g.,][]{BRYDEN+99}.
Planets may also explain the lower---by up to an order of
magnitude---accretion rates inferred for transition disks relative to
classical T Tauri stars \citep{NSM07,FM08,KIM+13}.  Since transition
disk masses are similar to those of T Tauri stars, lower accretion
rates onto the star suggest some of the accretion flow is diverted in
the inner regions \citep{NSM07}.  Numerical models of forming planets
corroborate the idea that protoplanets can take up a large fraction of
the circumstellar accretion flow \citep[e.g.,][]{VARNIERE+06b}.
Protoplanets may open gaps in less evolved disks as well, as
illustrated by the gapped-disk structure seen in HL Tau with ALMA;
multiple, low-mass protoplanets can potentially explain such
observations \citep[e.g.,][]{DZW14}.
%Indeed, such accretion onto
%a forming planet is required for formation before the disk disappears.
%Taking a typical value of $\dot{M}$ for T Tauri stars of $10^{-7}$
%M$_{\odot}$ yr$^{-1}$, and assuming just 1\% of the accretion flow is
%intercepted by the planet, I find $\dot{M}_{\rm planet} \sim 10^{-9}$
%  M$_{\odot}$ yr$^{-1}$.  This is sufficient to form a Jupiter-mass
%  planet in $\sim 10^6$ yr, within observed disk dissipation
%  timescales.

Accreting protoplanets may generate substantial luminosity
 \citep[e.g.,][]{PN05,ZHU14}:
\begin{equation}
L_{\rm acc,p} \sim \frac{G M_{\rm p} \dot{M_{\rm p}}}{R_{\rm p}}.
\label{eq:lacc}
\end{equation}
Here the $p$ subscripts indicate quantities related to the
protoplanet.
For planets of Jupiter mass and radius, and accretion rates of
$\dot{M}_{\rm p} \sim 10^{-9}$ M$_{\odot}$ yr$^{-1}$ ($\sim 10\%$ of
typical accretion rates onto T Tauri stars), 
$L_{\rm acc,p} >10^{-4}$ L$_{\odot}$.  However protoplanets may be
much larger than $R_{\rm J}$ at early times, in which case a given
$\dot{M}$ would produce less luminosity.

%An accreting planet
%is cooler than the star, and the observed contrast at longer
%wavelengths would be substantially greater than $10^{-4}$
%\citep[e.g.,][]{ZHU14}.    Such contrast can be achieved with current
%imaging techniques, and accreting protoplanets may be
%detectable  in current and upcoming imaging surveys.

Here I estimate the appearance of accreting protoplanets during
different evolutionary stages, assuming formation via core accretion.
\citet{ZHU14} recently computed synthetic SEDs for circumplanetary
accretion disks.  I argue that such models are only appropriate
after protoplanet atmospheres have undergone hydrodynamic collapse,
and I constrain the lifetime over which this evolutionary phase may be
observed.  I discuss the potential of current observational surveys to
detect accreting protoplanets at a range of evolutionary stages.

\begin{figure*}
\plotone{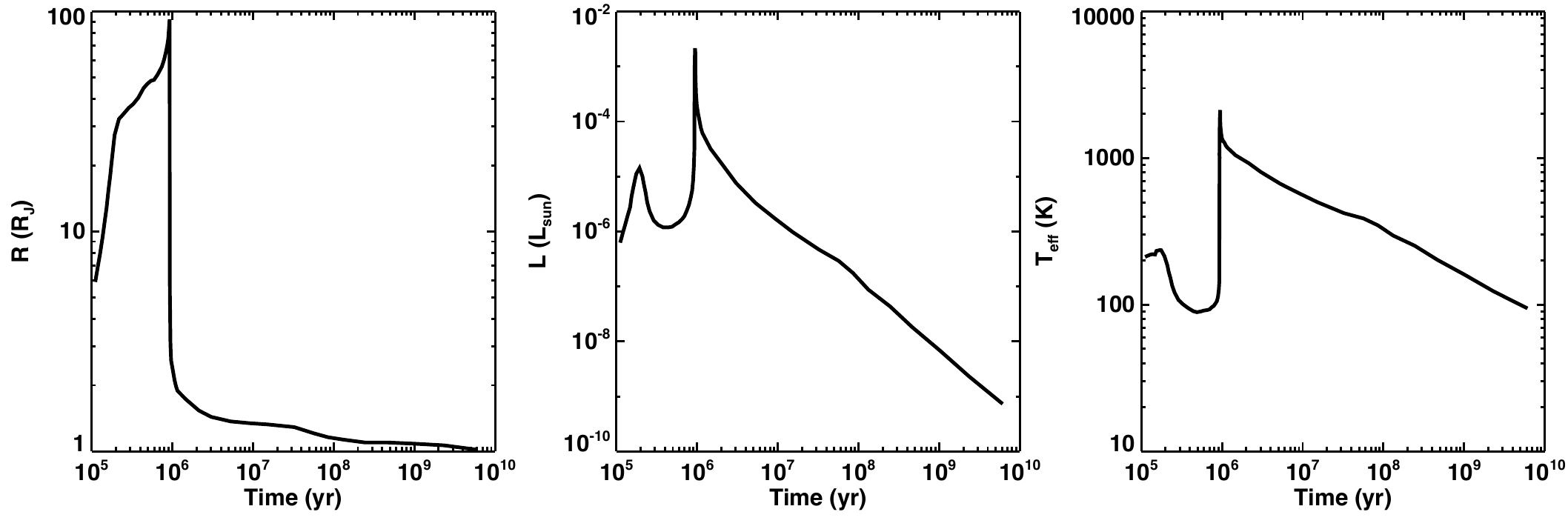}
\caption{Radii, luminosities, and effective temperatures for a giant
  planet undergoing gas accretion to eventually reach M$_{\rm J}$.  
  $R$ and $L$ are taken from \citet{MORDASINI+12}, and I used
  these to calculate $T_{\rm eff}$.
\label{fig:rlt}}
\end{figure*}

\section{Accretion onto Protoplanets}
\label{sec:acc}

Planet accretion begins with the assembly of a solid core.
If cores attain sufficient mass \citep[$\ga 10$ M$_{\oplus}$;
e.g.,][]{POLLACK+96}, the gravity of the
protoplanet is sufficient to overcome the tidal gravity of the central
star in a substantial ``feeding zone'',  roughly the size of the Hill
sphere:
\begin{equation}
 R_{\rm H} = a \left(\frac{M_{\rm p}}{3 M_{\ast}}\right)^{1/3}.
\end{equation}
As planets accrete mass, the Hill radius grows, and for a Jupiter-mass
planet at 5 AU around a solar-type star, 
$R_{\rm H} \approx 750$ R$_{\rm J}$ ($\approx 0.35$ AU).  

The scale height of circumstellar disks (at a few AU) is similar to or
larger than the size of the Hill sphere for forming gas giants (see
Section \ref{sec:gaps}),
leading to  spherical accretion onto protoplanets.
At early times protoplanet envelopes are in
hydrostatic equilibrium, with energy liberated by infalling
planetesimals providing pressure support against gravity.  
The high opacities in protoplanet
atmospheres result in hydrostatic envelopes filling a large fraction
of the Hill sphere \citep[e.g.,][]{POLLACK+96}.   
Because $R_{\rm H} > 100$ R$_{\rm J}$, gas does not accrete deep
into the potential well of the protoplanet. Although the
mass accretion rate of planetesimals is about an order of magnitude
lower than the gas accretion rate during this hydrostatic stage, the
luminosity is dominated by planetesimals, since these penetrate deeper
into the protoplanet envelope \citep{POLLACK+96}.  

Protoplanets in hydrostatic equilibrium accrete gas
slowly, since they must cool and contract in order to allow more
matter into the Hill sphere.  However once the mass of the gaseous
envelope reaches the ``crossover mass'' ($M_{\rm env} \approx
M_{\rm core} \approx 10$--20 M$_{\oplus}$), 
hydrostatic equilibrium breaks down and the
atmosphere collapses \citep[e.g.,][]{POLLACK+96}.  
Numerical simulations suggest that after hydrodynamic collapse, a
circumplanetary disk is formed within a fraction of the Hill radius
\citep[e.g.,][]{AB12}.   At this stage, the circumplanetary disk can
bring matter down close to $R_{\rm J}$, deep in the potential well of
the protoplanet, and high accretion luminosities are possible 
\cite[e.g., Equation \ref{eq:lacc};][]{PN05}.

\citet{POLLACK+96} predict that accreting protoplanets will have
hydrostatically supported envelopes for $\sim 7$ Myr.
This timescale is uncomfortably long compared to observed
lifetimes of gaseous circumstellar disks \citep[e.g.,][]{FEDELE+10}.
Planetary migration can accelerate the
accretion process, since the feeding zone continually takes in new
matter \citep{ALIBERT+05}.   Revised opacities for forming giant
planet atmospheres also result in substantially accelerated accretion,
because energy can be radiated away more efficiently 
\citep[e.g.,][]{MOVSHOVITZ+10}.
In these cases, the envelope reaches the
crossover mass---after which hydrodynamic collapse leads to runaway
gas accretion---in approximately a tenth the time required in the
\citet{POLLACK+96} models.  

After circumplanetary disk formation, the accretion rate onto the
protoplanet is limited by how fast the circumstellar disk
can supply material.
%, although $\dot{M}_{\rm p}$ may briefly reach
%$\sim 10^{-4}$ M$_{\rm J}$ yr$^{-1}$ while the protoplanet accretes
%the local reservoir of gas immediately after hydrodynamic collapse 
%\citep[e.g.,][]{MORDASINI+12}.   
%Observations and models of transition
%disks suggest that a large
%percentage---perhaps more than 90\%---of accretion through the
%circumstellar disk may be diverted
%to forming protoplanets \citep[e.g.,][]{NSM07,VARNIERE+06b}.  
%Taking $10^{-8}$ M$_{\odot}$ yr$^{-1}$ as a typical accretion rate for
%transition disks, 10--90\% of the accretion flow would produce $\dot{M}
%\sim 10^{-5}$--$10^{-6}$ M$_{\rm J}$ yr$^{-1}$.
Circumstellar accretion rates decline on $\sim 10^6$ yr timescales 
\citep[e.g.,][]{FEDELE+10}, suggesting that high accretion rates onto
protoplanets---which do not undergo hydrodynamic collapse until $\sim
10^6$ yr---are not likely to continue long.   Furthermore, any objects
that continued to accrete at such rates for $\ga 1$ Myr would 
exceed the planetary mass range.
%Known transition disks have inferred disk
%masses comparable to high-mass T Tauri disks,
%which suggests that they can sustain circumstellar disk accretion
%rates of $\sim 10^{-8}$ M$_{\odot}$ yr$^{-1}$.  
%$\dot{M}_{\rm p}$ during
%runaway accretion requires a large fraction of this accretion flow.
The observed accretion rates onto the central stars in
transition disk systems are $\la 10^{-9}$
M$_{\odot}$ yr$^{-1}$, which may indicate a large fraction of
the accretion flow is intercepted by protoplanets with $\dot{M}_{\rm
  p}$ near the runaway accretion rate
\citep[e.g.,][]{NSM07,VARNIERE+06b}.  
However, the stellar accretion rates can also be naturally explained with MRI-driven viscosity
\citep{CM07}.   It is thus unclear how much protoplanetary accretion
can be currently sustained in known transition disks.

\section{Gap Opening}
\label{sec:gaps}

Protoplanets whose Hill radii are comparable to the scale height of
the circumstellar disk can open gaps via dynamical processes
\citep[e.g.,][]{LP93}.  Such gaps are presumably linked with the
appearance of transition disks, and also render protoplanets easier to
observe.

To determine when protoplanets might open gaps, 
I first estimate the circumstellar disk scale
height.  For a flared disk around a typical T Tauri star, 
$H/R \sim 0.05$--0.1 within $\sim 10$ AU \citep[e.g.,][]{CG97}.
However, disk shadowing could lead to substantially smaller scale
heights \citep[e.g.,][]{DDN01}.  A lower bound can be estimated by
assuming that the disk is heated only by accretion.  For a constant
$\dot{M} = 10^{-8}$ M$_{\odot}$ yr$^{-1}$, the disk temperature is
$<100$ K beyond 1 AU, the sound speed is $<0.6$ km s$^{-1}$ (assuming
the gas is mostly H$_2$) and $H/R \sim c_{\rm s}/v_{\rm K} \la 0.025$
within $\sim 10$ AU.
%This is somewhat smaller than values typically
%assumed for the protosolar nebula \citep[$\sim 0.05$--0.1;
%e.g.,][]{LP93}.  This is because the accretion rate I assumed, while
%typical for the transition disk phase, is lower than the one used for
%determining the initial conditions in the protosolar nebula.

Before hydrodynamic collapse, the total mass of a protoplanet is $\la 20$
M$_{\oplus}$ \citep[e.g.,][]{POLLACK+96,MORDASINI+12}, and thus the
Hill radius is $\la 0.03 a$.   This is likely smaller than the
circumstellar disk scale height, and so the protoplanet is still
embedded in the disk at this stage.  Indeed, the protoplanet is
``attached'' to the disk, with continuous pressure and temperature
across the disk/protoplanet boundary \citep[e.g.,][]{MORDASINI+12}.
Accretion luminosity generated by hydrostatic protoplanets would thus
be attenuated by surface layers of the circumstellar disk.

When hydrodynamic collapse begins, the protoplanet envelope detaches
from the circumstellar disk, and forms a circumplanetary disk
\citep[e.g.,][]{AB12,PN05}.  
%It is around this time that the total
%protoplanet mass, $\ga 20$ M$_{\oplus}$, becomes sufficient for gap
%opening.  
Near the beginning of the runaway accretion phase, the protoplanet
mass is $\sim 90$ M$_{\oplus}$ \citep{MORDASINI+12},
and $R_{\rm H} \sim 0.05a$,
probably sufficient for gap opening. 

While protoplanets may open gaps in disks,
making cleared regions as large as those observed in transition disks
is difficult from a theoretical perspective.  Multiple planets have
been invoked to explain large clearings \citep{ZHU+11,DS11}, but such
models can not reproduce both the depleted regions and the observed
accretion rates onto the central stars.  Dust filtration through 
local pressure maxima created by even small disk gaps may also help to
deplete the dust population in large clearings
\citep[e.g.,][]{RICE+06,ZHU+12}, although this mechanism fails to
clear out small dust grains. Radiation pressure, either from the central
star \citep[e.g.,][]{ECH06} or from the accreting protoplanet itself
\citep{OWEN14} may help to clear out the small dust.  

Since several different physical mechanisms may be responsible for
creating or maintaining the cleared regions in transition disks, one
should be cautious about the exact criteria used to determine when
protoplanets open gaps.  For example, 
any non-monotonic pressure profile associated
with a protoplanet may provide a seed for other physical mechanisms
to clear a gap.  It is thus difficult to be certain that hydrostatic
protoplanets---especially as they grow in mass and luminosity at later
times---will be completely enshrouded by circumstellar disk matter.
Furthermore, if lower-mass protoplanets do not open deep gaps like
those seen
in transition disks, they may open smaller gaps.  Evidence for
this is seen in the HL Tau disk, where the gapped disk structure
observed with ALMA can be explained
with multiple, $\sim 60$ M$_{\oplus}$ protoplanets \citep{DZW14}.

\section{Protoplanet SEDs Over Time}
\label{sec:seds}
To predict the appearance of accreting protoplanets, I
use calculations of the envelope structure of a forming giant planet
\citep[e.g.,][]{POLLACK+96,MORDASINI+12}.  I consider
the first 10 Myr or so, because known transition disks have typical
ages of a few Myr, with the oldest at $\sim 10$ Myr.  I adopt the
models of \citet{MORDASINI+12}, which
predict that runaway accretion occurs after $\sim 0.95$ Myr.  While longer
timescales for the onset of runaway accretion are possible
\citep[e.g.,][]{POLLACK+96}, the exact timescale does not affect the
results for pre- or post-runaway accretion stages.

\begin{figure}
\plotone{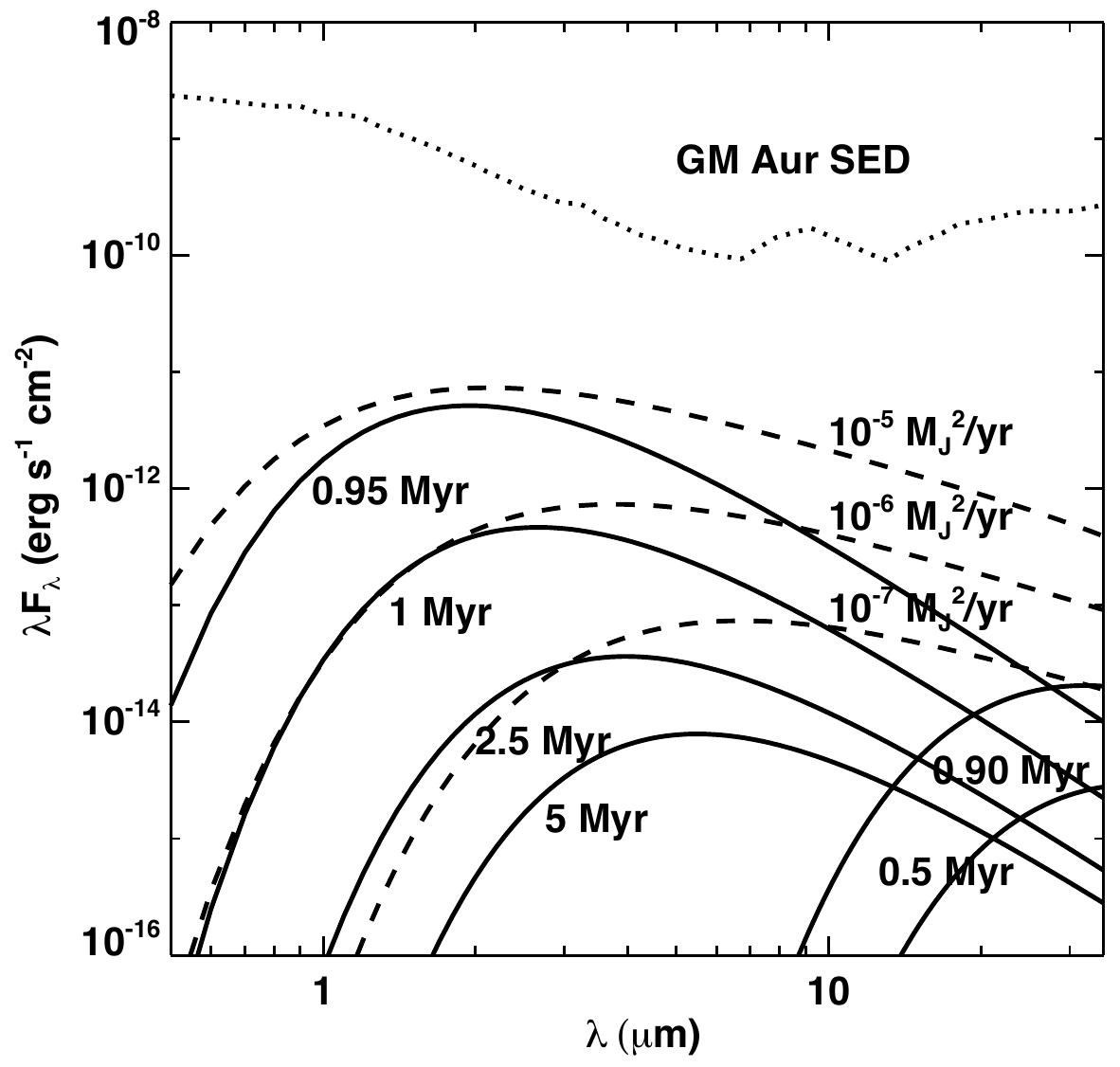}
\caption{SEDs calculated for 
%radii and effective temperatures of
  the accreting protoplanet
  shown in Figure \ref{fig:rlt} ({\it solid curves}). Several illustrative
  timesteps are shown, with particular focus on the  hydrodynamic
  collapse of the atmosphere at $\sim 0.95$ Myr.  These models assume
  that all luminosity is radiated from the protoplanetary
  surface.  Models where the luminosity is generated in
  circumplanetary disks are shown with dashed curves, labeled by the
  product of planet mass and accretion rate.
  I also
  include the observed SED of the transition disk GM Aur
  \citep[adjusted to a distance of 100 pc; adapted from][]{ZHU14}, to
  demonstrate the contrast between
  protoplanet and central star+circumstellar disk. 
\label{fig:accplanets}}
\end{figure}

Using planetary radii and accretion luminosities calculated as a function of
time \citep{MORDASINI+12}, I determine effective temperatures (Figure
\ref{fig:rlt}).    Radii and effective temperatures are then used to
generate blackbody SEDs as a function of time. All models are at an
assumed distance of 100 pc. 

Before hydrodynamic collapse, $R_{\rm p} \approx R_{\rm H}/3 \approx 100$
R$_{\rm J}$ \citep[e.g.,][]{MORDASINI+12,LISSAUER+09}.  This large
radius, combined with a modest accretion luminosity of 
$\sim 10^{-6}$ L$_{\odot}$, produces cool spectra, with peak
wavelengths $>10$ $\mu$m (Figure \ref{fig:accplanets}).  Since these
protoplanets are likely still attached to circumstellar disks, their
spectra will probably suffer additional extinction from matter in the
circumstellar disk surface layers.

After hydrodynamic collapse, when protoplanets detach from
circumstellar disks and probably open gaps, $R_{\rm p}$ approaches
$R_{\rm J}$, the accretion rate increases, and the luminosity climbs
to $\sim 10^{-3}$ L$_{\odot}$.  This leads to a bright protoplanet
with $T_{\rm eff} > 10^3$ K (Figure \ref{fig:accplanets}).  The models
of \citet{MORDASINI+12} assume that runaway accretion occurs for $\sim
10^5$ years, until the planet reaches a Jupiter-mass, after which the
circumstellar disk disappears and accretion ceases.  After this, 
the luminosity declines from $\sim
10^{-3}$ L$_{\odot}$ to $\sim 10^{-6}$ L$_{\odot}$, and $T_{\rm eff}$ declines
from $\sim 10^3$ to $\sim 600$ K, within $\sim 10$
Myr,  similar to  ``hot-start'' planetary atmosphere models
\citep[e.g.,][]{SB12}.  

These calculations assume that all luminosity is radiated from the
protoplanetary radius.  However hydrodynamic collapse likely forms a
circumplanetary disk \citep{PN05,AB12}, which would produce accretion
luminosity from somewhat larger radii.  \citet{ZHU14} recently
computed SEDs for circumplanetary disks with accretion rates similar
to those expected for runaway accretion.  I reproduce similar models
here using a simplified treatment.  Considering a disk extending from
2 $R_{\rm J}$ to $R_{\rm H}/3$, I compute
blackbody SEDs for a series of annuli and sum them.  SEDs for disks
around a Jupiter-mass planet with $\dot{M}_{\rm p}$ between $10^{-5}$
and $10^{-7}$ M$_{\rm J}$ yr$^{-1}$ are shown in Figure
\ref{fig:accplanets}.  These curves are nearly identical to those in
\citet{ZHU14}.

A circumplanetary accretion disk with $\dot{M} M_{\rm p} = 10^{-5}$
M$_{\rm J}^2$ yr$^{-1}$ produces a similar luminosity to the
calculated SED of a protoplanet undergoing runaway accretion.
The main difference between the two models is that the disk model
reprocesses light at longer wavelengths, leading to a slower falloff
in flux \citep[see also][]{ZHU14}.  As protoplanets cool, their
luminosities decline (see curves for $\ge 1$ Myr in Figure
\ref{fig:accplanets}).  If circumstellar and circumplanetary accretion
continues at these later times, but perhaps declines from its peak
rate, then the SEDs may manifest additional emission, particularly at
longer wavelengths (see the $10^{-6}$ and 10$^{-7}$ M$_{\rm J}^2$
yr$^{-1}$ curves in Figure \ref{fig:accplanets}).  

Whether accretion luminosity is generated at protoplanetary surfaces
or in accretion disks, the infrared emission is compact.  For $\lambda
<5$ $\mu$m, any observed
emission would be more compact than $\sim 0.01$ AU.  At wavelengths up
to 30 $\mu$m, emission may be distributed on scales several times
larger.

\section{Comparison with Observations}
Contrast ratios between modeled protoplanets (Figure
\ref{fig:accplanets})
and a typical transition disk (GM Aur)
are listed in Table \ref{tab:contrast}.  Contrasts 
could be lower for less evolved gapped-disks like HL Tau, which exhibit
stronger near-IR excess emission than transition disks.  However this
additional near-IR emission would be less important for protoplanets
at large stellocentric radii.

Extreme adaptive optics on large telescopes can reach
contrasts of $\sim 10^{-6}$ at $1''$ separations
\citep[e.g.,][]{SKEMER+14}, but such scales are
larger than typical transition disk holes.
%: at 100 pc, 100 AU subtends $1''$.  
Within $0\rlap{.}''1$ (10 AU at 100 pc)
%; appropriate for transition disks), 
achieved contrasts are closer to $10^{-3}$ \citep[e.g.,][Sallum et
al. 2015]{KI12,BILLER+14}.  Thus, $10^{-3}$ is a reasonable benchmark
against which to compare the computed contrast ratios of 
protoplanet models.

\begin{deluxetable*}{lcccccc}
\tabletypesize{\scriptsize}
\tablewidth{0pt}
\tablecaption{Protoplanet/Transition Disk Contrast Ratios
\label{tab:contrast}}
\tablehead{\colhead{Model} & \colhead{2 $\mu$m} & \colhead{4 $\mu$m}
  & \colhead{5 $\mu$m} & \colhead{10 $\mu$m} & \colhead{20 $\mu$m} & 
\colhead{30 $\mu$m}}
\startdata
{\bf 0.5 Myr} & $<10^{-29}$ & $<10^{-17}$ & $<10^{-13}$ & $<10^{-8}$ & $<10^{-6}$ & $<10^{-5}$ \\
{\bf 0.9 Myr} & $<10^{-22}$ & $<10^{-12}$ & $<10^{-10}$ & $<10^{-6}$ & $<10^{-4}$ & $<10^{-4}$ \\
{\bf 0.95 Myr} & $10^{-2}$ & $10^{-2}$ & $10^{-2}$ & $10^{-3}$ & $10^{-4}$ & $10^{-4}$ \\
$ $ $\dot{M} M_{\rm p} = $ $10^{-5}$ & $10^{-2}$ & $10^{-2}$ & $10^{-2}$ & $10^{-2}$ & $10^{-3}$ & $10^{-3}$ \\
{\bf 1 Myr} & $10^{-3}$ &  $10^{-3}$ & $10^{-3}$ & $10^{-4}$ & $10^{-4}$ & $10^{-5}$ \\
$ $ $\dot{M} M_{\rm p} = $$10^{-6}$ & $10^{-3}$ & $10^{-3}$ & $10^{-2}$ & $10^{-3}$ & $10^{-3}$ & $10^{-4}$ \\
{\bf 2.5 Myr} & $10^{-5}$ & $10^{-4}$ & $10^{-4}$ & $10^{-4}$ & $10^{-5}$ & $10^{-6}$ \\
$ $ $\dot{M} M_{\rm p} = $$10^{-7}$ & $10^{-4}$ & $10^{-4}$ &
$10^{-3}$ & $10^{-4}$ & $10^{-4}$ & $10^{-4}$ \\
{\bf 5 Myr} & $10^{-6}$ & $10^{-5}$ & $10^{-4}$ & $10^{-5}$ & $10^{-5}$ & $10^{-6}$ \\
%$10^{-4}$ & $10^{-1}$ & $10^{-1}$ & $10^{-1}$ & $10^{-1}$ & $10^{-2}$ & $10^{-2}$ \\
\enddata
\tablecomments{Models in bold assume hydrodynamic collapse and runaway
  accretion around 0.95 Myr, and then $\dot{M}=0$ for $>1$ Myr. For
  comparison, I also include circumplanetary accretion disk models
  with $\dot{M}$ declining from the peak runaway rate.
   Since
  protoplanets may still be embedded in circumstellar disks at $<0.95$
  Myr, the contrasts are listed as upper limits for such objects. }
 % of M$_{\rm J}^2$ yr$^{-1}$) listed in
 % Figure \ref{fig:accdisk}.}
\end{deluxetable*}

Detecting hydrostatic protoplanets is extremely difficult
at the expected contrasts.  
If emission from a protoplanet is reprocessed
by the atmosphere of the circumstellar disk (Section \ref{sec:gaps}), 
the contrasts at infrared wavelengths may even be more challenging than
implied by Table \ref{tab:contrast}.
The best hope for detecting hydrostatic protoplanets is observing at
long IR wavelengths\footnote{Contrasts are higher in the (sub-)mm due to
  the luminous long-wavelength SEDs of transition disks.}, and
targeting transition disks with particularly large clearings (where
observations can achieve higher contrasts).

Protoplanets in, or shortly after, the runaway accretion
phase are more easily detectable.    
%Contrasts approaching $10^{-3}$
%at separations smaller than $0\rlap{.}''1$ can now be achieved 
%at $\sim 2$--5 $\mu$m wavelengths.   This can be compared with
Modeled protoplanets with $\dot{M}$ near the runaway accretion
rate of $10^{-5}$ M$_{\rm J}$ yr$^{-1}$ all have contrasts of $\sim
10^{-2}$ with the transition disk SED at some wavelength between 2 and
5 $\mu$m (Table \ref{tab:contrast}).   After runaway
accretion as the circumstellar and circumplanetary accretion rates
decline, protoplanets may fall below detection thresholds.

The most promising candidate is LkCa 15, where
localized structure at both $K$ and $L$ bands is detected and inferred
to orbit around the central star over time \citep{KI12}. The contrasts
with the central star (plus its transition disk) are $\sim 2 \times
10^{-3}$ at $K$ and $\sim 7 \times 10^{-3}$ ar $L$.
Similar observations of T Cha suggest the presence of some
structure at $K$ and $L$ bands within the cleared region of the disk, 
with contrasts of $\sim 5 \times 10^{-3}$ at $K$ and $\sim 8 \times
10^{-3}$ at $L$ \citep[][Sallum et al. 2015]{HUELAMO+11}.   
These contrasts, for both sources, are similar to 
expectations for a protoplanet that has recently undergone runaway
accretion (Table \ref{tab:contrast}).

However images of the emission within the cleared regions of both the
LkCa 15 and T Cha transitions disks show complex structure that does
not appear to trace compact sources.  The maximum outer
radius of circumplanetary disks is $\sim R_{\rm H}/3 \approx 0.5$
AU for even a 10 Jupiter-mass planet within 10 AU of a sun-like star.
Moreover, infrared emission is generated in the inner regions
of circumplanetary disks ($\la 0.01$ AU).
In contrast, extended structure is seen on $\sim 5$ AU scales in LkCa 15
\citep{KI12}, and even larger scales in the T Cha disk (Sallum \&
Eisner in prep).  
%Such extended structures are a problem for (single)
%accreting protoplanet models.  
While multiple protoplanets might
explain the observed structures (and may also help explain the origin
of the large cleared inner region in the LkCa 15 disk), disk-planet
interactions may also
lead to extended, asymmetric structure in transition disks
\citep[e.g.,][]{LYRA+09,FGM10}.  Future observations at higher
resolution are needed to elucidate the nature of the emission.

Coronographic imaging observations suggest a companion in the HD
169142 transition disk, with a detection at only $L$ band, and
non-detections at $J,H$, and $K$ \citep{REGGIANI+14,BILLER+14}.
This may be a case where the protoplanet's contrast with the
transition disk SED only exceeds the detection threshold at certain
wavelengths.  For example, the $\dot{M} M_{\rm p} = 10^{-6}$ M$_{\rm
      J}^2$ yr$^{-1}$ model has  contrast of $\sim 10^{-3}$ at $K$
band, but 4--6 $\times 10^{-3}$ at $L$ and $M$ bands.  The inferred
contrast of the companion is $\sim 3 \times 10^{-3}$ at $L$ band,
matched reasonably well with the model prediction \citep[see
also][]{ZHU14}.

A spatially extended ($\ga 12$ AU diameter)  companion has been
claimed in the HD 100546 disk at a stellocentric radius of $\sim 70$
AU \citep{QUANZ+13,CURRIE+14}.   \citet{QUANZ+14} confirm this
companion at both $L$ and $M$ bands, and argue for a compact
source surrounded by extended emission.  The compact object has
a contrast of $\sim 2 \times 10^{-4}$ at both $L$ and $M$ bands
\citep{QUANZ+14}, consistent with a cooling protoplanet at
$\sim 2.5$ Myr, or a circumplanetary disk with $\dot{M}
M_{\rm p} \sim 10^{-7}$ M$_{\rm J}^2$ yr$^{-1}$.  However the putative
protoplanet appears to be still embedded in the circumstellar disk,
suggesting a younger evolutionary state.

Finally, mid-IR imaging has suggested the presence of self-luminous
companions in two other transition disks, SR 21 \citep{EISNER+09b} and
TW Hya \citep{ARNOLD+12}.  The posited companion in
the SR 21 disk is too red and luminous compared to the models
presented here, suggesting it is unlikely to be an accreting protoplanet.
The companion suggested around TW Hya has a
luminosity similar to the $\dot{M} M_{\rm p} = 10^{-6}$ M$_{\rm
  J}^2$ yr$^{-1}$ model, but a redder spectral shape.  If this
companion is real, the accretion would have to fall in only to
$\sim 10$ R$_{\rm J}$ to explain the color.   Such a large inner disk
radius could result from a highly magnetized protoplanet, whose
magnetosphere would truncate the disk.  Magnetospheric accretion in
this putative protoplanet could also lead to high H$\alpha$ luminosity
\citep{ZHU14}, potentially detectable in follow-up observations at
visible wavelengths \citep[e.g.,][]{CLOSE+14}.

The models presented here can be used to estimate the
probability that accreting proptoplanets would be detected in any
transition disk. 
For current detection thresholds,
Jupiter-mass protoplanets are detectable during the runaway accretion
phase and for $\sim 0.5$ Myr after, as they cool
(Table \ref{tab:contrast}).    Higher mass planets stay in the
runaway accretion stage longer and start cooling from a higher
temperature: a 10 M$_{\rm J}$ planet might remain detectable for $\sim
1$ Myr.   Taking the circumstellar disk lifetime to
be 3 Myr \citep[e.g.,][]{FEDELE+10}, one expects $\sim 15$--30\% of
disks to harbor bright protoplanets.    This is similar to the percentage of
transition disks around young stars $\sim 10$--30\%
\citep[e.g.,][]{MUZEROLLE+10,ANDREWS+11}.  
Thus, it seems likely that a large
fraction of these transition disks would host a protoplanet with a
high accretion luminosity.

While many transition disks
have no detected planets despite observations with the required contrast
\citep[e.g.,][]{KRAUS+11,EVANS+12}, observations probed a
limited semimajor axis range.  Most candidate protoplanets have
semimajor axes $\ga 10$ AU, 
located near $\lambda/D$ for the telescopes used in their detections.
As larger telescopes become available, smaller
separations can be probed, and more accreting protoplanets may be found.

\section{Summary}
Protoplanets that are still hydrostatically supported and attached to
their circumstellar disks are faint, red, and may be extincted by
circumstellar disk surface layers.   Detecting such objects is
extremely difficult, although observations at long-IR wavelengths are
the best hope.  In contrast, protoplanets whose atmospheres have
undergone hydrodynamic collapse may be detectable during their runaway
accretion phase.  Whether accretion luminosity is assumed to be
generated at the surface of the protoplanet or in a circumplanetary
disk, such objects appear relatively bright compared to the central
stars and circumstellar disks.  Their infrared emission should be
compact, at odds with most claimed observations of protoplanetary
companions.  While the lifetime of the runaway
accretion phase is short compared to circumstellar disk lifetimes, I
argue that protoplanets should be detectable in
a large fraction of transition disks, and that as larger telescopes
probe smaller separations additional protoplanets will be found.

$ $
\\
This work was supported by NSF AAG grant 1211329, and benefitted from
discussions with Phil Armitage.

\end{document}